\documentclass[a4paper,twoside]{article}
\newcommand{\affil}[1]{$^{\rm #1}$}
\usepackage[authoryear]{natbib}
\bibpunct{(}{)}{;}{a}{}{,}
\usepackage{graphicx}
\date{} %Please leave the date blank
\newcommand{\aap}{Astronom. and Astrophys.}
\newcommand{\aaps}{Astronom. and Astrophys. Suppl. Ser.}
\newcommand{\aj}{Astronom. J.}
\newcommand{\apjs}{Astrophys. J. Suppl.}
\newcommand{\araa}{Annu. Rev. Astronom. Astrophys.}
\newcommand{\mnras}{Monthly Notices Roy. Astronom. Soc.}
\def\fm{\hbox{$.\!\!^m$}}
\def\fs{\hbox{$.\!\!^s$}}
\def\degr{\hbox{$^\circ$}}

\title{\large\bf\flushleft Physical Parameters of the Visually Close Binary Systems Hip70973 and Hip72479}
\author{\parbox{\textwidth}{\flushleft
\vspace{-0.5cm}
{\it Mashhoor Al-Wardat\affil{A,B} }\\
\vspace{0.4cm}
{\small \affil{A}\, Physics Department, Yarmouk University, P.O.B. 566 Irbid, 21163 Jordan }\\
{\small \affil{B}\, Department of Physics, Al-Hussein Bin Talal University, P.O.Box 20,
71111, Ma'an, Jordan. Email: mwardat@ahu.edu.jo}}}
\begin{document}
\twocolumn[
\begin{changemargin}{.8cm}{.5cm}
\begin{minipage}{.9\textwidth}
\vspace{-1cm}
\maketitle
%
%
%%%%%%%%%%%%%     ABSTRACT    %%%%%%%%%%%%%
%Abstract of no more than 200 words here.
\small{\bf Abstract:} Atmospheric modelling of the components of the visually close  binary
systems Hip70973 and Hip72479 was used to estimate the individual physical parameters of their components. The model atmospheres  were constructed using a grid of  Kurucz solar metalicity blanketed models, and used to compute a synthetic spectral energy distribution for each component separately, and hence for the combined system. The total observational spectral energy distributions of the systems were used as a reference for the comparison with the synthetic ones. We used the feedback modified parameters and iteration method to get the best fit between synthetic and  observational spectral energy distributions.
 The physical parameters of the
components of the system Hip70973 were derived as: $T_{\rm eff}^{a}
=5700\pm75$\,K, $T_{\rm eff}^{b} =5400\pm75$\,K, log $g_{a}=4.50\pm0.05$,
 log $g_{b}=4.50\pm0.05$, $R_{a}=0.98\pm0.07 R_\odot$,  $R_{b}=0.89\pm0.07 R_\odot$, and $\pi=26.25 \pm 1.95 $ mas, with G4 \& G9 spectral types.
And  those of the system Hip72479 as: $T_{\rm eff}^{a}
=5400\pm50$\,K, $T_{\rm eff}^{b} =5180\pm50$\,K, log $g_{a}=4.50\pm0.05$,
 log $g_{b}=4.60\pm0.05$, $R_{a}=0.89\pm0.07 R_\odot$,  $R_{b}=0.80\pm0.07 R_\odot$, and $\pi=23.59 \pm 1.00 $ mas with G9 \& K1 spectral types.

%%%%%%%%%%%%%     KEYWORDS    %%%%%%%%%%%%%
\medskip{\bf Keywords:} stars: physical parameters, binaries, visually close binary systems, atmospheres modelling, Hip70973, Hip72479
% Please write all keywords in lower case. PASA uses the
% standard list of subject headings adopted by The Astrophysical Journal
% and available from http://www.journals.uchicago.edu/ApJ/keywords_text.html.
% Keywords are separated by em-dashes, i.e. ---

%%%%%%%%DO NOT EDIT%%%%%%%%%%%%
\medskip
\medskip
\end{minipage}
\end{changemargin}
]
\small
%%%%%%%%EDIT FROM HERE%%%%%%%%%%%%

\section{Introduction}

The Hipparcos mission revealed that many previously known   single stars were actually binary or multiple systems \citep{1998AstL...24..673S, 2002A&A...385...87B}. Most  of these resolved systems are nearby stars that appear as a single star even with the largest ground-based telescopes except when we use high resolution techniques like speckle interferometry  (SI)\citep{2002A&A...385...87B, 2010AJ....139..743T} and adaptive optics (AO) \citep{2011MNRAS.413.1200R, 2005AJ....130.2262R}. These systems are known as visually close binary systems (VCBS).

The study of binary systems plays an important role in determining several key stellar parameters, which is more complicated in the case of VCBS.
Hundreds of binary systems with periods on the order of 10 years or less, are routinely observed with high resolution techniques. In spite of that, there is still a paucity  of  individual physical parameters for the systems'
components. So, spectrophotometry with atmospheric modelling is a complementary solution to this problem, by giving an
accurate determination of the effective temperature, radius and luminosity  for each component of a binary
system. The method was successfully applied to some binary system
like ADS11061, Cou1289, Cou1291, Hip11352 and Hip11253  \citep{2002BSAO...53...51A, 2007AN....328...63A, 2009AN....330..385A, 2009AstBu..64..365A}.

 The two binary systems Hip70973 and Hip72479 are  well known VCBS.  So, they fulfil the requirements to be analyzed by the aforementioned method in order to get their complete physical parameters. Table~\ref{table1} contains basic data of the systems from SIMBAD, NASA/IPAC and The Geneva-Copenhagen survey of the Solar neighborhood \citep{2004A&A...418..989N}. Table~\ref{table2}
contains data from Hipparcos and Tycho Catalogues \citep{1997yCat.1239....0E}.

 The system Hip70973 was discovered by Rossiter (1938.51) with the 27 inch (0.69 m) telescope at the Lamont-Hussey Observatory \citep{2000AJ....119.2422D}. Orbits of the system had been calculated
by \cite{1960JO.....43...13C}, \cite{1970A&AS....1..115M}, \cite{1981ApJS...45..559H} (two orbits; the first one with period 45.4 yr and dynamical parallax  $0.''020$,  and the second one with period 22.4 yr and dynamical parallax  $0.''029$),  \cite{1999A&A...341..121S} and \cite{2000AJ....119.2422D} (elements of this orbit are listed in Table ~\ref{orb}).

The system Hip72479 (ADS9397) was discovered by Aitken in 1916.40 at the Lick Observatory. Its orbits had been calculated by \cite{1954CiUO..114Q.236V, 1945MNSSA...4....3V, 1964ROCi..123...62V}, \cite{1965AJ.....70...19E, 1967ARA&A...5..105E} (different orbits using photometrical parallax  $0.''026$), \cite{1999A&A...341..121S} and \cite{2000AJ....119.2422D} (elements of this orbit are listed in Table ~\ref{orb}).

  The estimated parameters will enhance our knowledge about stellar parameters in general, and consequently help  in understanding the formation
and evolution mechanisms of binary stellar systems.

\begin{table}[!h]
\begin{center}
\caption{Basic data of the systems} \label{table1}
\begin{tabular}{lccc}\hline
  & Hip70973  & Hip72479& ref.  \\
   &RST4529 & A2983 &\\
 \hline
$\alpha_{2000}$ & $14^h 31^m 00\fs650$ &  $14^h 49^m 13\fs621$ &1\\
$\delta_{2000}$&$-05\degr48' 08.''46$ &$+10\degr12' 52.''06$ &1\\
WDS & 14310-0548 &14492+1013 &1\\
 Tyc &  4996-131-1 & 921-918-1 &1\\
 HD &  127352  & 130669 &1\\
 Sp. Typ. & G5 & K2V &1\\
 E(B-V) &0.0499& 0.0268 &2\\
 $A_v$&$0\fm156$&$0\fm083$&2\\
  $\log T_{\rm eff}$ & 3.722 & 3.705& 3 \\
 $[Fe/H]$ &$ -0.14$ & 0.11&3 \\
  $M_ v $ & $4\fm75$ &  $5\fm32$ &3 \\
    \hline
\end{tabular}
\\
$^1${SIMBAD},
$^2${NASA/IPAC:http://irsa.ipac.caltech.edu},
$^3${\cite{2004A&A...418..989N}}.
\end{center}
\end{table}

\begin{table}[!h]
\begin{center}
\caption{Data from Hipparcos and Tycho Catalogues} \label{table2}
\begin{tabular}{lcc}\hline
  & Hip70973   & Hip72479 \\
 & HD127352  & HD130669 \\
 \hline
  $V_J(Hip)$ & $7\fm68$ & $8\fm42$  \\
  $B_T$ & $8\fm667\pm0.014$ & $9\fm540\pm0.020$ \\
 $V_T$ & $7\fm781\pm0.011$ & $8\fm534\pm0.014$ \\
 $(B-V)_J(Tyc)$ & $0\fm775\pm0.003$ & $0\fm866\pm0.007$ \\
 $\pi_{Hip}$ (mas) & $26.04\pm1.04$ & $24.21\pm1.29$\\
 $\pi_{Tyc}$ (mas) & $30.6\pm8.7$ & $20.8\pm10.6$\\
 $\pi_{Hip}^*$ (mas) & $24.31\pm0.89$ & $22.59\pm1.23$\\

\hline
\end{tabular}\\
$^*${Reanalyzed Hipparcos parallax \cite{2007A&A...474..653V}}
\end{center}
\end{table}

\begin{table}[!h]
\begin{center}
\caption{Orbital elements of the systems \citep{2000AJ....119.2422D}} \label{orb}
\begin{tabular}{lcc}\hline
 Hip & 70973   & 72479 \\
WDS & 14310-0548  & 14492+1013 \\
 \hline
  $P$ (yr) & $22.98\pm0.30$ & $9.98\pm0.04$  \\
 $T^* $    & $1993.62\pm0,02$ & $1988.059\pm0.03$  \\
 $e  $   & $0.499\pm0.010$& $0.491\pm0.001$ \\
 $a$ (arcsec)& $0.243\pm0.002$& $0.127\pm0.001$ \\
 $i$ (deg) & $49.1\pm2.0 $ & $45.8\pm2.0$    \\
$ \Omega$ (deg) & $13.8\pm2.0$  & $142.3\pm2.0$ \\
$\omega$ (deg) & $121.0\pm2.5$ &$156.8\pm3.0 $\\
$\pi_{dyn}$ (mas) & $23.2$ & $21.1$\\
\hline
\end{tabular}
$^*${Periastron transit time (yr).}
\end{center}
\end{table}

\section{Atmospheric modelling}
\subsection{Hip70973}

 We adopted the magnitude difference between the two components $\triangle m=0\fm56$ as the average of all $\triangle m$ measurements under the speckle filters  $550nm/40$ \& $551nm/22$ (see Table ~\ref{deltam1}) as the closest filters to the visual. This value was used as an input to the equation:
 \begin{eqnarray}
\label{eq1}
\frac{f_1}{f_2}=2.512^{-\triangle m},
\end{eqnarray}
\noindent
along with the visual magnitude of the combined system $m_v=7\fm68 $ from Table~\ref{table2} as an input to the equation:
 \begin{eqnarray}
\label{eq2}
m_v=-2.5\log(f_1+f_2).
\end{eqnarray}

From these we  calculated a preliminary individual $m_v$ for each
component as: $m_{va}=8\fm19$ and $m_{vb}=8\fm75$.

\begin{table}[!ht]
\begin{center}
\caption{Magnitude difference between the components of the
system Hip70973, along with filters used to obtain the observations. }
\label{deltam1}
\begin{tabular}{lcc}
\noalign{\smallskip}
\hline
\noalign{\smallskip}
   $\triangle m $& filter ($\lambda/\Delta\lambda$)& ref.  \\
\hline
\noalign{\smallskip}
 $0\fm42\pm0.15$ & $V_{Hp}: 550nm/40 $& 1 \\
 $0\fm52$ & $550nm/40 $& 2 \\
 $0\fm55$ & $698nm/39$ & 2\\
 $0\fm60$ & $551nm/22 $  &3\\
 $0\fm60$ & $657nm/5 $  &3\\
 \hline
\noalign{\smallskip}
\end{tabular}
\\
$^1${\cite{1997yCat.1239....0E}},
$^2${\cite{2008AJ....136..312H}},
$^3${\cite{2010AJ....139..743T}}.
\end{center}
\end{table}

%\begin{table}[!ht]
%\begin{center}
%\caption{total synthetic Johnson, Str\"{o}mgren and Tycho  magnitudes and  colour indices of both systems %\cite{2002BSAO...53...58A, 2008AstBu..63..361A}.}
%\label{obsmag}
%\begin{tabular}{lcc}
%\noalign{\smallskip}
%\hline
%\noalign{\smallskip}
%  & Hip70973 &  Hip72479 \\
% \hline
%\noalign{\smallskip}
% $B_J$ & $8\fm46 \pm 0.06  $ & $9\fm31 \pm 0.06  $\\
% $V_J$ & $7\fm63 \pm 0.06 $ & $8\fm38 \pm 0.06 $\\
% $R_J$ & $7\fm21 \pm 0.07 $  & $7\fm88 \pm 0.07 $\\
% $(B-V)_J$ & $0\fm84 \pm 0.08$  & $0\fm95 \pm 0.08$\\
% $v$ & $8\fm91 \pm 0.06$ & $9\fm84 \pm 0.06$ \\
% $b$& $8\fm14 \pm 0.06$ & $8\fm94 \pm 0.06$\\
% $y$ & $7\fm59 \pm 0.06$ & $8\fm32 \pm 0.06$ \\
%$v-b$& $0\fm77 \pm 0.08$ & $0\fm91 \pm 0.08$\\
% $b-y$ & $0\fm56 \pm 0.08$  & $0\fm62 \pm 0.08$ \\
%  $B_T$ & $8\fm68 \pm 0.06  $ & $9\fm56 \pm 0.06  $ \\
% $V_T$ & $7\fm73 \pm 0.06 $ & $8\fm49 \pm 0.06 $ \\
% $(B-V)_T$ & $0\fm95 \pm 0.08$ & $1\fm07 \pm 0.08$\\
%\noalign{\smallskip}
%\hline
%\end{tabular}
%\end{center}
%\end{table}

\begin{table}[!ht]
\begin{center}
\caption{Magnitude difference between the components of the
system Hip72479, along with filter used to obtain the observations. }
\label{deltam2}
\begin{tabular}{lcc}
\noalign{\smallskip}
\hline
\noalign{\smallskip}
   $\triangle m $& filter ($\lambda/\Delta\lambda$)& ref.  \\
\hline
\noalign{\smallskip}
 $0\fm09\pm0.72$ & $V_{Hp}: 550nm/40 $& 1 \\
  $1\fm00$ &$657nm/5 $  &2\\
  $0\fm40$ & $551nm/22 $  &2\\
  $0\fm70$ &$657nm/5 $  &2\\
  $0\fm60$ & $551nm/22 $  &2\\
 \hline
\noalign{\smallskip}
\end{tabular}
\\
$^1${\cite{1997yCat.1239....0E}},
$^2${\cite{2010AJ....139..743T}}.
\end{center}
\end{table}

%The aforementioned observational SED and SIMBAD spectral type of the system show that it is a solar type main sequence star. So,  we can relay on
Using the following main sequence relations and tables \citep[e.g.,][]{1992adps.book.....L, 2005oasp.book.....G}:
\begin{eqnarray}
\label{eq3}
M_v=m_v+5-5\log(d)-A,\\
\label{eq4}
\log(R/R_\odot)= 0.5 \log(L/L_\odot)-2\log(T/T_\odot),\\
\label{eq5}
\log g = \log(M/M_\odot)- 2\log(R/R_\odot) + 4.43,
\end{eqnarray}
we  calculated the preliminary input  parameters (bolometric magnitudes, luminosities and effective temperatures) of the individual components.
We used bolometric corrections of
\cite{1992adps.book.....L} \& \cite{2005oasp.book.....G},  $T_\odot=5777 \rm{K}$ and extinction ($A_v$) given in Table~\ref{table1} by NASA/ IPAC.

These  calculated input parameters allow construction of  model
atmospheres for each component using  grids of  Kurucz's 1994
blanketed models (ATLAS9), where we used solar abundance model
atmospheres. Hence a spectral energy distribution for each component can be built.

The total energy flux from a binary star is created from the net
luminosity of the components $a$ and $b$ located at a distance $d
$ from the Earth. So we can write:

\begin{eqnarray}
\label{eq6}
   F_\lambda \cdot d^2 = H_\lambda ^a \cdot R_{a} ^2 + H_\lambda ^b
\cdot R_{b} ^2,
\end{eqnarray}
 \noindent from which

\begin{eqnarray}
\label{eq7}
 F_\lambda  = (R_{a} ^2/d)^2(H_\lambda ^a + H_\lambda ^b \cdot(R_{b}/R_{a})^2) ,
\end{eqnarray}
\noindent
 where $H_\lambda ^a $ and  $H_\lambda ^b$ are the fluxes from a unit
surface of the corresponding component. $F_\lambda$ here
represents the total SED of the system.

%Here we used the system's parallax  measurment by Hipparcos
%($\pi=26.04\pm1.04,d=38.40 pc$) as a temporary (subject to change) input for the calculations.

Within the criteria of the best fit, which are the maximum values
of the absolute flux, the shape of the continuum, and the profiles of
the absorption lines, and starting with the preliminary calculated parameters,
 many attempts were made to achieve the best
fit between the observed flux and the total computed one using the
iteration method of different sets of parameters.

Using Hipparcos modified parallax $(\pi=24.31 \pm 0.89 \textrm{mas. Table ~\ref{table2}})$, the best fit  was achieved using the following set of parameters:

  $$ T_{\rm eff}^{a}=5700\pm75{\rm K}, T_{\rm eff}^{b} =5400\pm75{\rm K},$$ $$ \log g_{a}=4.50\pm0.05 , \log g_{b}=4.50\pm0.05,$$
  $$R_{a}=1.05\pm0.07R_\odot \,\, \textrm{and} \,\,R_{b}=0.96\pm0.07R_\odot .$$

 But the values of the estimated radii disagree with those given by \cite{2005oasp.book.....G}, \cite{1992adps.book.....L}  and the R-L-T relation (equation ~\ref{eq4}) for the main sequence stars. According to equation ~\ref{eq7}, this disagreement refers to a misestimation in the parallax of the system, which means that changing the parallax of the system affects  the values of the components' radii.

\begin{figure}[h]
\includegraphics[angle=0,width=8.5cm]{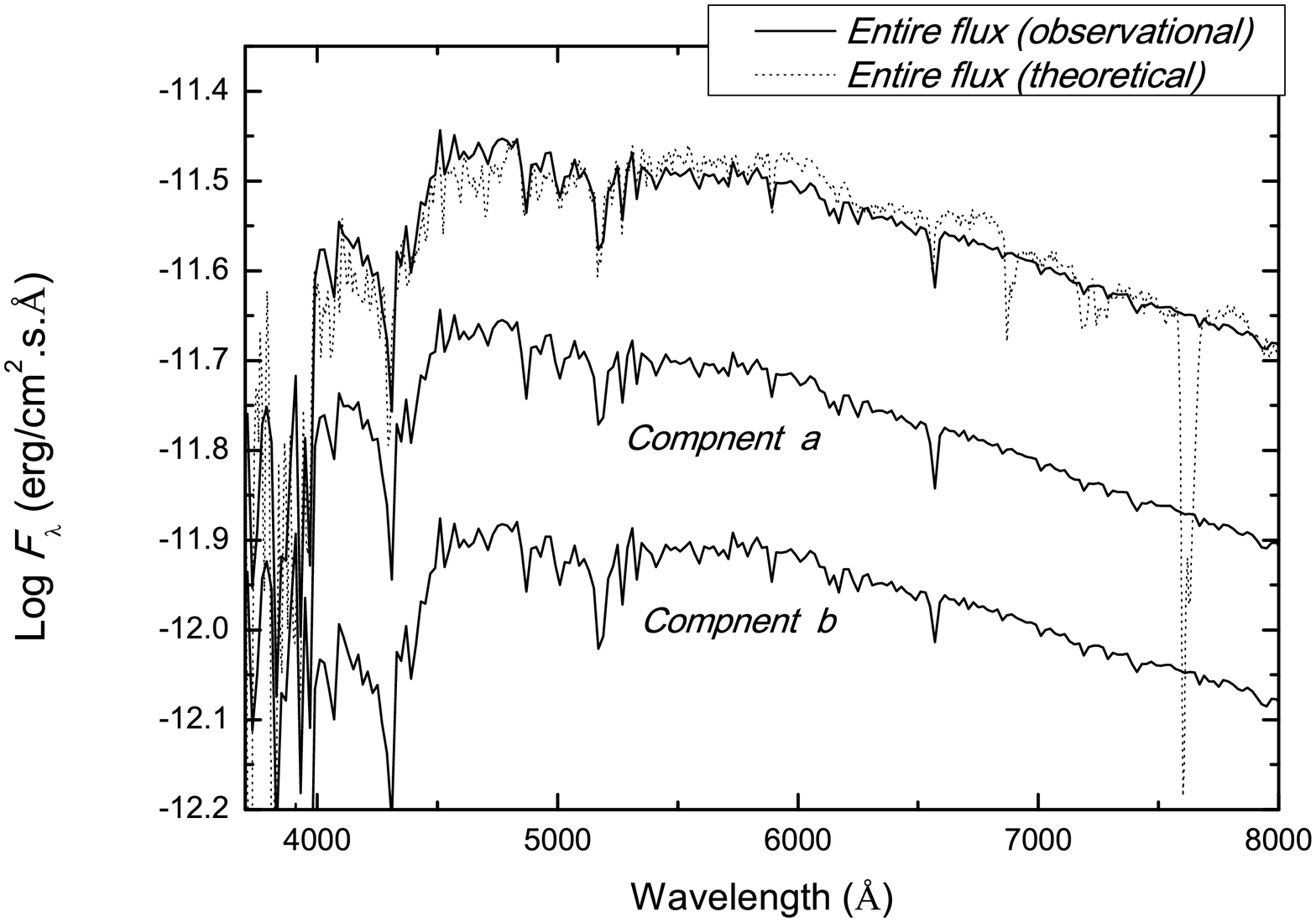}
\caption{Dotted line: the total observational SED in the continuous spectrum of the
 system Hip70973. Solid lines: the  total computed SED of the two components with $d=38.10\pm3.04$\,pc ($\pi=26.25 \pm 1.95 $ mas),
  the computed flux of the primary component with $T_{\rm eff}=5700\pm75$\,K,
 log $g=4.50\pm0.05, R=0.98\pm0.07R_\odot$, and the computed flux of the secondary
component with
 $T_{\rm eff} =5400\pm75$\,K, log $g=4.50\pm0.05, R=0.89\pm0.07 R_\odot $.}
\label{hip70973}
\end{figure}

  So, in order to reach reliable parameters for the system, we went the other way; i.e. we started with the radii which are compatible with the tables of \cite{2005oasp.book.....G} and changed the parallax. Keeping in mind the values of the total observational $V_J, B_T, V_T$ and $\bigtriangleup m$ as our goal in achieving the best fit between the synthetic and observational total absolute fluxes, we reached that using the following set of parameters (Fig.~\ref{hip70973}):
   $$ T_{\rm eff}^{a}=5700\pm75{\rm K}, T_{\rm eff}^{b} =5400\pm75{\rm K},$$ $$ \log g_{a}=4.50\pm0.05 , \log g_{b}=4.50\pm0.05,$$
   $$R_{a}=0.98\pm0.07R_\odot, R_{b}=0.89\pm0.07R_\odot ,$$
      and $$d=38.10\pm3.04\,\textrm{pc} (\pi=26.25 \pm 1.95  \textrm{mas.}),$$

Thus the  luminosities of the components follow as:
$L_a=0.91\pm0.08 L_\odot$,  and $L_b=0.61\pm0.05 L_\odot$. These values represent adequately  enough  the parameters of the systems' components.

Fig.~\ref{hip70973} shows the best fit between the total synthetic SED and the  observational one  taken from \cite{2002BSAO...53...58A}.
Note that some of the strong lines and depressions, especially in
the red part of the spectrum (around $\lambda 6867\textrm{\AA}$, $\lambda
7200\textrm{\AA}$, and $\lambda 7605\textrm{\AA}$), are $\rm H_2O$ and $\rm O_2$
telluric lines and depressions.

Depending on the tables of \cite{2005oasp.book.....G} or using \cite{1992adps.book.....L} $Sp-T_{\rm
eff}$ empirical relation,  the spectral types  of the system's components can be estimated as G4 \& G9.

\subsection{Hip72479}

 For this system we adopted the magnitude difference between the two components $\triangle m=0\fm50$ which is the average of all $\triangle m$ measurements under the speckle filters  $551nm/22$ (see Table ~\ref{deltam2}) as the closest filters to the visual. This value when used as an input to equation~\ref{eq1}, along with the total visual  magnitude of the system $m_{v}=8\fm534$ from Table~\ref{table2} as an input to equation~\ref{eq2}, results in the preliminary individual $m_v$ for each of the
component as: $m_{va}=9\fm07$ and $m_{vb}=9\fm57$.

\begin{figure}[!h]
\includegraphics[angle=0,width=8.5cm]{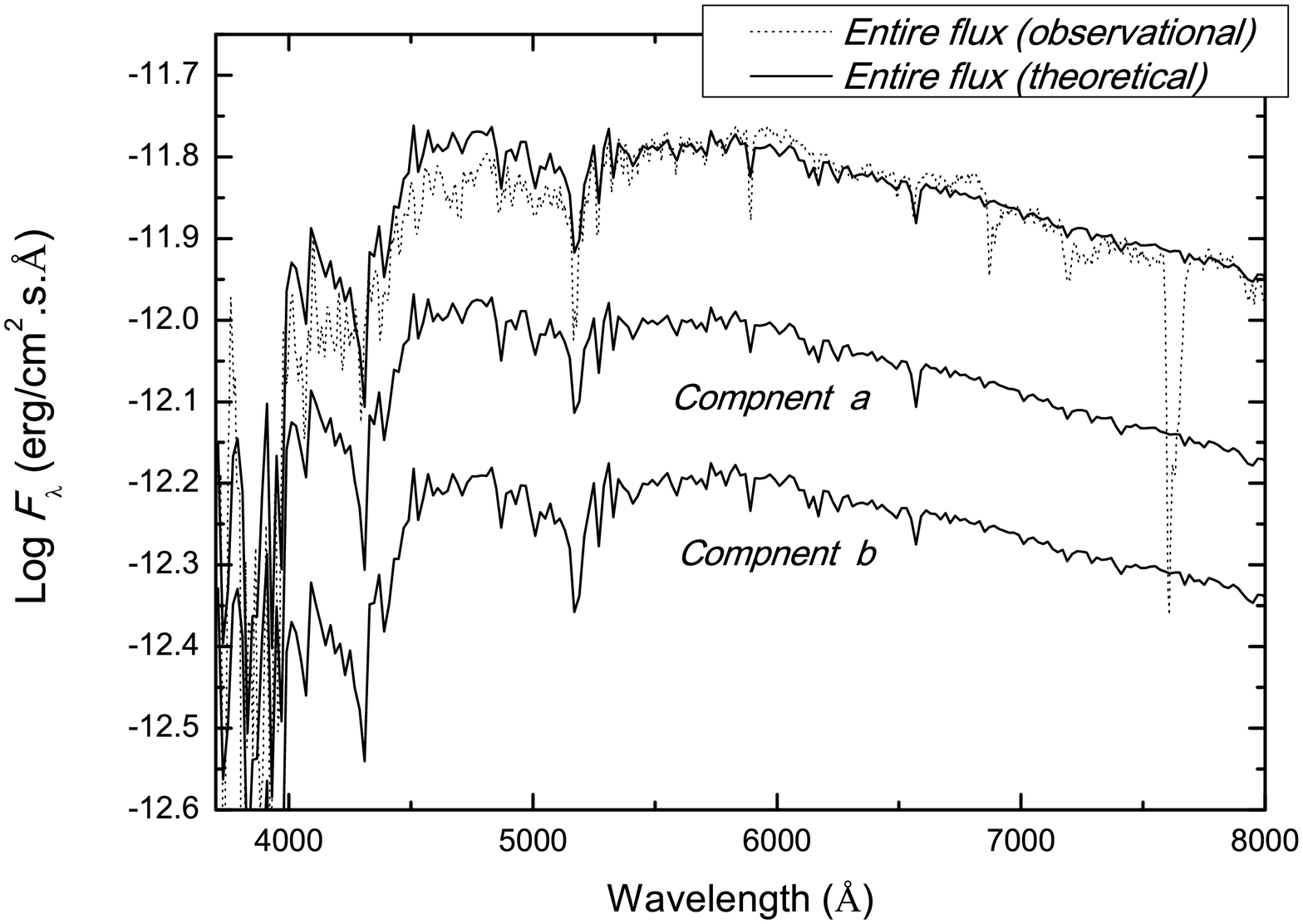}
\caption{Dotted line: the total observational SED in the continuous spectrum of the
 system Hip72479. Solid lines: the  total computed SED of the two components with $d=42.40\pm1.85$\,pc. ($\pi=23.59 \pm 1.00 $ mas),
  the computed flux of the primary component with $T_{\rm eff}=5400\pm50$\,K,
 log $g=4.50\pm0.05, R=0.89\pm0.07R_\odot$, and the computed flux of the secondary
component with
 $T_{\rm eff} =5180\pm50$\,K, log $g=4.60\pm0.05, R=0.80\pm0.07 R_\odot $.}
\label{hip72479}
\end{figure}

Following the same procedures explained in the previous section, and using Hipparcos modified parallax  measurement  as initial input
($\pi=22.59\pm1.23, d=44.26$pc) as a temporary  input for the calculations,
  the best fit between the total synthetic SED and the  observational one  taken from \cite{2002BSAO...53...58A} was achieved using the following
set of parameters (Fig. ~\ref{hip72479}):

  $$ T_{\rm eff}^{a}=5400\pm50{\rm K}, T_{\rm eff}^{b} =5180\pm50{\rm K},$$ $$ \log g_{a}=4.50\pm0.05 , \log g_{b}=4.60\pm0.05,$$
  $$R_{a}=0.92\pm0.07R_\odot \,\, \textrm{and} \,\,R_{b}=0.84\pm0.07R_\odot .$$

Again here,  the values of the estimated radii do not fit exactly those given by \cite{2005oasp.book.....G}, \cite{1992adps.book.....L}  and the R-L-T relation (equation ~\ref{eq4}) for the main sequence stars. So, we recalculated the absolute flux using \cite{2005oasp.book.....G} radii as postulate values leaving the parallax subject to change.
 Following the guidance of the total observational $V_J, B_T, V_T$ and $\bigtriangleup m$, the best fit between the synthetic and observational total absolute fluxes was achieved  using the following set of parameters
$$ T_{\rm eff}^{a}=5400\pm50{\rm K}, T_{\rm eff}^{b} =5180\pm50{\rm K},$$ $$ \log g_{a}=4.50\pm0.05 , \log g_{b}=4.60\pm0.05,$$
  $$R_{a}=0.89\pm0.07R_\odot,  R_{b}=0.80\pm0.07R_\odot,$$ and $$d=42.40\pm1.85\,\textrm{pc} (\pi=23.59 \pm 1.00  \textrm{mas.}),$$

Thus the luminosities follow as:
$L_a=0.61\pm0.05 L_\odot$,  and $L_b=0.41\pm0.03 L_\odot$,
\noindent
with spectral types  G9 \& K1 for the primary and secondary components respectively.

\begin{table}[!ht]
\begin{center}
\caption{Magnitude difference between the components of the
system Hip72479, along with filter used to obtain the observations. }
\label{deltam2}
\begin{tabular}{lcc}
\noalign{\smallskip}
\hline
\noalign{\smallskip}
   $\triangle m $& filter ($\lambda/\Delta\lambda$)& ref.  \\
\hline
\noalign{\smallskip}
 $0\fm09\pm0.72$ & $V_{Hp}: 550nm/40 $& 1 \\
  $1\fm00$ &$657nm/5 $  &2\\
  $0\fm40$ & $551nm/22 $  &2\\
  $0\fm70$ &$657nm/5 $  &2\\
  $0\fm60$ & $551nm/22 $  &2\\
 \hline
\noalign{\smallskip}
\end{tabular}
\\
$^1${\cite{1997yCat.1239....0E}},
$^2${\cite{2010AJ....139..743T}}.
\end{center}
\end{table}

%\begin{table}[!h]
%\begin{center}
%\caption{The different parallaxes and relevant radii of the best fit between synthetic and observational absolute fluxes for the system Hip72479.}
%\label{radii2}
%\begin{tabular}{lccc}\hline
%  \multicolumn{2}{c}{ $\pi$ (mas)}& R1 & R2 \\

% \hline

% Hip                   & $24.21\pm1.29$ & $0.96\pm0.08$         & $0.94\pm0.08$ \\
%  ${\textrm{Hip}}^1$   & $22.59\pm1.23$ & $1.02\pm0.08$         & $0.94\pm0.08$\\
% Tyc                   & $20.8\pm10.6$  & $1.10\pm0.13$         & $1.02\pm0.13$  \\
% $\textrm{Egg}^2_{dyn}$&   26           &  $0.90\pm0.13 $       &   $0.82\pm0.13$  \\
% $\textrm{Doc}^3_{dyn}$&  21.1          &    $1.10\pm0.13 $     &   $1.01\pm0.13$   \\
% \hline
%\end{tabular}\\
%$^1${Reanalyzed Hipparcos parallax \cite{2007A&A...474..653V}},
%$^2$\cite{1965AJ.....70...19E, 1967ARA&A...5..105E},
%$^3$\citep{2000AJ....119.2422D}.
%\end{center}
%\end{table}

\section{Synthetic photometry}
In addition to the direct comparison, we can check the reliability of our method of estimating the physical and geometrical parameters  by comparing the observed  magnitudes of the combined system from different ground or space based telescopes  with the synthetic ones. For that, we used the following relation \citep{{2006AJ....131.1184M},{2007ASPC..364..227M}}:
\begin{equation}
m_p[F_{\lambda,s}(\lambda)] = -2.5 \log \frac{\int P_{p}(\lambda)F_{\lambda,s}(\lambda)\lambda{\rm d}\lambda}{\int P_{p}(\lambda)F_{\lambda,r}(\lambda)\lambda{\rm d}\lambda}+ {\rm ZP}_p\,,
\end{equation}
to calculate the total  and individual synthetic magnitudes of the systems, where $m_p$ is the synthetic magnitude of the passband $p$, $P_p(\lambda)$ is the dimensionless sensitivity function of the passband $p$, $F_{\lambda,s}(\lambda)$ is the synthetic SED of the object and $F_{\lambda,r}(\lambda)$ is the SED of the reference star (Vega).  Zero points (ZP$_p$) from  \cite{2007ASPC..364..227M} (and references there in) were adopted.

 The results of the calculated magnitudes and colour  indices of the combined system and individual components, in different photometrical systems,  are shown in Tables~\ref{synth1}~\& \ref{synth2}.

\begin{table}[!ht]
\small
\begin{center}
\caption{ Magnitudes and colour indices  of the synthetic spectra of the  system Hip70973.}
\label{synth1}
\begin{tabular}{lcccc}
\noalign{\smallskip}
\hline
\noalign{\smallskip}
Sys. & Fil. & total & comp.& comp.\\
     &     & $\sigma=\pm0.03$&   a    &     b      \\
\hline
\noalign{\smallskip}
Joh-  & $U$          & 8.74 & 9.17 & 9.96 \\
 Cou.          & $B$ & 8.43   &  8.92 &  9.55  \\
               & $V$ & 7.68 &  8.20 &  8.73 \\
               & $R$ & 7.28 &  7.82 & 8.29  \\
               &$U-B$& 0.31 & 0.25 & 0.41 \\
               &$B-V$&0.76  &  0.72 &  0.82 \\
               &$V-R$& 0.41 &  0.38 & 0.44 \\
  \hline
\noalign{\smallskip}
  Str\"{o}m.        & $u$ & 9.88 & 10.31 &  11.11  \\
                    & $v$ & 8.84 & 9.30  & 9.99  \\
                    & $b$ & 8.09 & 8.59 &  9.17 \\
                    &  $y$& 7.65 & 8.17 &  8.69  \\
                    &$u-v$& 1.04 &1.01& 1.12 \\
                    &$v-b$& 0.75 & 0.71 & 0.82 \\
                    &$b-y$& 0.45 & 0.43& 0.48 \\
  \hline
\noalign{\smallskip}
  Tycho       &$B_T$  & 8.63   & 9.10 & 9.77   \\
              &$V_T$  & 7.76   &8.28 & 8.82  \\
              &$B_T-V_T$& 0.87 & 0.83& 0.95\\
\hline
\noalign{\smallskip}
\end{tabular}
\end{center}
\end{table}

% Hip72479

\begin{table}[!ht]
\small
\begin{center}
\caption{ Magnitudes and colour indices  of the synthetic spectra of the  system Hip72479.}
\label{synth2}
\begin{tabular}{lcccc}
\noalign{\smallskip}
\hline
\noalign{\smallskip}
Sys. & Fil. & total & comp.& comp.\\
     &     & $\sigma=\pm0.03$&   a    &     b      \\
\hline
\noalign{\smallskip}
Joh-            & $U$ & $9.74$ & $10.19$ & $10.91$ \\
 Cou.          & $B$ & 9.28   &  9.78 &  10.36  \\
               & $V$ & 8.43 &  8.96 &  9.47 \\
               & $R$ & 7.97 &  8.52 & 8.98  \\
               &$U-B$& 0.46 & 0.41 & 0.56 \\
               &$B-V$&0.85  &  0.82 &  0.89 \\
               &$V-R$& 0.46 &  0.44 & 0.49 \\
  \hline
\noalign{\smallskip}
  Str\"{o}m.        & $u$ & 10.90 & 11.34 &  12.08  \\
                    & $v$ & 9.74 &  10.22  & 10.86  \\
                    & $b$ & 8.88 & 9.40 &  9.93  \\
                    &  $y$& 8.39 & 8.92 &  9.37  \\
                    &$u-v$& 1.15 &1.12 & 1.23 \\
                    &$v-b$& 0.86 & 0.82 & 0.92 \\
                    &$b-y$& 0.49 & 0.48& 0.51 \\
  \hline
\noalign{\smallskip}
  Tycho       &$B_T$  & 9.51   & 10.00 & 10.60   \\
              &$V_T$  & 8.52   &9.05 & 9.56  \\
              &$B_T-V_T$& 0.98 & 0.95& 1.04 \\
\hline
\noalign{\smallskip}
\end{tabular}
\end{center}
\end{table}

%\section {Formation and evolution of the systems}
\section {Results and discussion}

\begin{figure}[h]
\includegraphics[angle=0,width=8.5cm]{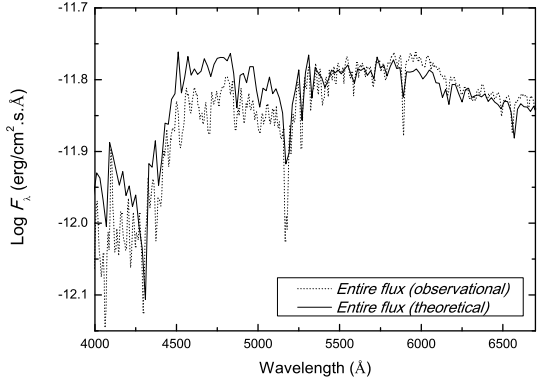}
\caption{An expanded look at the fit between the total synthetic SED's and the  observational ones for the left part of the spectrum of the system Hip72479.}
\label{hip72479blue}
\end{figure}

Looking deeply at the achieved best fit between the total synthetic SED's and the  observational ones (Figs. ~\ref{hip70973}~\&~\ref{hip72479}), we see that there is a good overall coincidence in the  maximum values of the absolute fluxes and the shape of the continuum except for the blue part of the spectrum around $\lambda 4700\textrm{\AA}$ (Fig.~\ref{hip72479blue}). A part of this disagreement is due to the lack of some opacities in the synthetic SED's and to the difference in the resolution between the synthetic and observational spectra.

 A comparison between the  synthetic magnitudes, colours and magnitude differences  with the observational ones (Table~\ref{tablecopmarison})  shows a very good consistency within the error values.  This gives a good indication for the reliability of the  estimated parameters of the individual components of the system, which are listed  in Tables~\ref{tablef1}~\&~\ref{tablef2}.

\begin{table}[!ht]
\small
\begin{center}
\caption{Comparison between the observational and synthetic  magnitudes, colours and magnitude differences for both systems.}
\label{tablecopmarison}
\begin{tabular}{lcc}
\noalign{\smallskip}
\hline
\noalign{\smallskip}
      Hip70973.      & $\textrm{Obs}.^*$    & Synth. (this work)  \\
\hline
\noalign{\smallskip}

  $V_J$     & $7\fm68$          & $7\fm68\pm0.03$  \\
  $B_T$     & $8\fm667\pm0.014$ & $8\fm63\pm0.03$ \\
  $V_T$     & $7\fm781\pm0.011$ & $7\fm76\pm0.03$ \\
  $(B-V)_J$ & $0\fm775\pm0.003$ & $0\fm76\pm0.04$ \\
  $\triangle m $& $0\fm56^\dag$ & $0\fm53\pm0.04$\\

\hline
     Hip72479 & &\\
\hline
  $V_J$     & $8\fm42$          & $8\fm43\pm0.03$  \\
  $B_T$     & $9\fm540\pm0.020$ & $9\fm51\pm0.03$ \\
  $V_T$     & $8\fm534\pm0.014$ & $8\fm52\pm0.03$ \\
  $(B-V)_J$ & $0\fm866\pm0.007$ & $0\fm85\pm0.04$ \\
  $\triangle m $& $0\fm50^\dag$        & $0\fm51\pm0.04$\\
\hline

\noalign{\smallskip}
\end{tabular}\\
$^*${See Table~\ref{table2}}.\\
$^\dag${Average value for the filters  $550nm/40$ \& $551nm/22$ (Tables ~\ref{deltam1}\,\,\&  ~\ref{deltam2}) }.
\end{center}
\end{table}

  Concerning  the accuracy of our estimated parallaxes,  it depends on the accuracy of the input radii  as it is clear from equation ~\ref{eq6} and ~\ref{eq7}. This means that the first solution which depends on Hipparcos new parallax is not excluded at all, especially if we know that the values of the radii were higher in the tables of \cite{2005oasp.book.....G} comparing with tables of \cite{1992adps.book.....L}.

\begin{figure}[!ht]
%\centerline{\psfig{figure=evol.eps,width=0.5\textwidth,clip=}}
\includegraphics[angle=0,width=8.5cm]{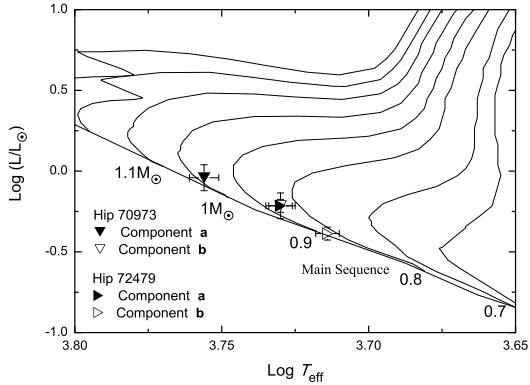}
 \caption{The  systems' components  on the evolutionary tracks (solar abundance) of  \cite{2000A&AS..141..371G}. }
 \label{evol2}
\end{figure}

 Fig.~\ref{evol2} shows the positions of the components on the
 evolutionary tracks of  \cite{2000A&AS..141..371G}, where the error bars in the figure include the effect of the parallax  uncertainty.
 The ages of the systems can be established from the evolutionary tracks.

 It is clear from the parameters of the system's components and their positions on the evolutionary tracks that they are  solar type main sequence stars, in the early stages of their life. Depending on the formation theories,   fragmentation
is a possible  process for the formation of the systems studied in this work.
Where \cite{1994MNRAS.269..837B} concludes that fragmentation of a rotating disk
around an incipient central protostar is possible, as long as
there is continuing infall.  \cite{2001IAUS..200.....Z} pointed out that
hierarchical  fragmentation during rotational collapse has been
invoked to produce binaries and multiple systems.

\begin{table}[!ht]
\small
\begin{center}
\caption{Parameters of the components of the system Hip70973.}
\label{tablef1}
\begin{tabular}{lcc}
\noalign{\smallskip}
\hline
\noalign{\smallskip}
Component & a &  b  \\
\hline
\noalign{\smallskip}
$T_{\rm eff}$\,(K) & $5700\pm75$ & $5400\pm75$ \\
Radius (R$_{\odot}$) & $0.98\pm0.07$ & $0.89\pm0.07$ \\
$\log g$ & $4.50\pm0.05$ & $4.50\pm0.05$ \\
$L (L_\odot)$ & $0.91\pm0.08 $  & $0.61\pm0.05$\\
$M_v^*$         &  $4.97 \pm0.10$ &     $5.77\pm0.10$ \\
Mass, ($M_{\odot})^*$& $1.07\pm0.08$ & $0.94\pm0.05$  \\
Sp. Type$^*$ & G4 &G9 \\
\hline
\multicolumn{1}{l}{Parallax (mas) }
    & \multicolumn{2}{|c}{$26.25 \pm 1.95 $}\\
  \multicolumn{1}{l}{Age (Gy) }
    & \multicolumn{2}{|c}{ $2.7\pm 0.3$}\\
\hline
\noalign{\smallskip}
\end{tabular}
$^*${depending on the tables of  \cite{2005oasp.book.....G}}.
\end{center}
\end{table}

\begin{table}[!ht]
\small
\begin{center}
\caption{Parameters of the components of the system Hip72479.}
\label{tablef2}
\begin{tabular}{lcc}
\noalign{\smallskip}
\hline
\noalign{\smallskip}
Component & a &  b  \\
\hline
\noalign{\smallskip}
$T_{\rm eff}$\,(K) & $5400\pm50$ & $5180\pm50$ \\
Radius (R$_{\odot}$) & $0.89\pm0.07$ & $0.80\pm0.07$ \\
$\log g$ & $4.50\pm0.05$ & $4.60\pm0.05$ \\
$L (L_\odot)$ & $0.61\pm0.05 $  & $0.41\pm0.03$\\
$M_v^*$         & $5.77\pm0.10$ &     $6.25\pm0.10$ \\
Mass, ($M_{\odot})^*$& $0.94\pm0.05$ & $0.85\pm0.04$  \\
Sp. Type$^*$ & G9 & K1 \\
\hline
\multicolumn{1}{l}{Parallax (mas) }
    & \multicolumn{2}{|c}{$23.59 \pm 1.00 $}\\
  \multicolumn{1}{l}{Age (Gy) }
    & \multicolumn{2}{|c}{ $2.7\pm 0.3$}\\
\hline
\noalign{\smallskip}
\end{tabular}
$^*${depending on the tables of  \cite{2005oasp.book.....G}}.
\end{center}
\end{table}

\section{Conclusions}

The analysis of the two VCBS, Hip70973 and Hip72479, using atmospheric modelling results in the  following main
conclusions.

\begin{enumerate}
    \item The parameters of the systems'  components were estimated
  depending on the best fit between the observational SED  and
  synthetic ones built using the atmospheric modelling of the individual components.
  \item New parallaxes of the systems  were estimated from these stellar parameters.
   \item From the parameters of the systems' components and their positions on the evolutionary tracks, we showed  that the components within each system are similar solar type main sequence stars  (G4~\&~G9 for Hip70973 and G9~\&~K1 for Hip72479).
      \item The total and individual $UBVR$ Johnson-Cousins, $uvby$ Str\"{o}mgren and $BV$ Tycho synthetic magnitudes and colours of the systems were calculated.
   \item Because of the high similarity of the two components within each system, fragmentation is proposed as the most
  likely process for the formation and evolution of both systems.
\end{enumerate}

\section*{Acknowledgments}
This work was done during the research visit to Max Planck Institute for Astrophysics-Garching, which was funded by DFG. The author thanks Professor Peter Cottrell for reviewing earlier versions of the manuscript.  This work made use of SAO/NASA, SIMBAD, IPAC data systems and CHORIZOS code of photometric and spectrophotometric data  analysis. The author expresses  the  sincere thanks to the critical comments from the anonymous referee  that greatly improved the quality of the paper.

%\bibliographystyle{aa}
%\bibliography{alwardat}

\end{document}